# High temperature interaction between molten $Ni_{50}Al_{50}$ alloy and $ZrB_2$ ultra-high temperature ceramics

Rafał Nowak[*], Grzegorz Bruzda, Wojciech Polkowski[*],

*Łukasiewicz Research Network - Krakow Institute of Technology, Zakopianska 73 Str., Krakow, Poland*
*Corresponding author email:
rafal.nowak@kit.lukasiewicz.gov.pl (R.N); wojciech.polkowski@kit.lukasiewicz.gov.pl (W.P.)*

**Abstract:** In this work, $Ni_{50}Al_{50}$ alloy is taken into consideration as potential brazing material for joining $ZrB_2$ ultra-high temperature ceramic. The results of experimental study on high temperature interfacial phenomena between molten binary $Ni_{50}Al_{50}$ alloy and polycrystalline $ZrB_2$, are shown. A sessile drop method combined with a capillary purification procedure was applied to investigate the wetting behavior of $Ni_{50}Al_{50}/ZrB_2$ couple during holding for 400 seconds at temperature of 1688°C (i.e. at $T=1.02T_m$). It was found that the molten $Ni_{50}Al_{50}$ rapidly wets and spreads over the surface of $ZrB_2$, while involved reactive infiltration into the solid substrate allowed reaching a final contact angle of ~0° in 250 sec. The wetting kinetics was much faster than that reported in the literature for Cu, Ag or Au tested at $T=1.05T_m$. The solidified couple was subjected to SEM/EDS microstructural characterization in order to reveal a course of interfacial phenomena. The results point towards (I) an existence of Ni-enriched Ni-Al/$ZrB_2$ surface interfacial layer; (II) a formation of infiltration zone assisted by reactively formed $Al_2O_3$ due to a reaction with Al-rich melt and (III) a partial transfer of $ZrB_2$ phase to $Ni_{50}Al_{50}$ alloy by a dissolution/precipitation mechanism.

1. Introduction

Transition metals borides (including TaB, $HfB_2$ or $ZrB_2$) with melting points above 3000°C have been identified as promising candidates for ultra-high applications requesting high thermal fluxes and severe surface stresses (e.g. in sharp leading-edge components) [**1**]. Among these ultra-high-temperature ceramics (UHTCs), a special attention is given to $ZrB_2$, due to its relatively low density as compared to other borides, high strength, hardness, thermal conductivity and corrosion resistance [2]. A crucial aspect regarding practical implementation of $ZrB_2$ ceramic and composites is their joining to similar (ceramic) or dissimilar (metallic) materials. A proper design and thermal management in ultra-high-temperature devices allow introducing metallic brazing materials to join components made of $ZrB_2$. A successful material candidate should possess a high melting point, excellent oxidation resistance and a good thermophysical/chemical compatibility (reflected e.g. by a good wetting and interfacial bonding) with the UHTC. In this regard, various metallic materials have been examined so far including various (Ag,Cu)-(Ti,Zr,Hf) based alloys [3], boron doped amorphous Ni-base brazes [4] (both having liquidus $T_L$~1050-1150°C) or 60Pd-40Ni and 65Pd-35Co based alloys ($T_m$~1219-1238°C) [5]. In terms of materials having higher melting points, Valenza et al. [6] have performed wettability experiments on $ZrB_2$ ceramics by pure Ni, Ni-17B at.%, and Ni-50B at.% at 1500 and 1200°C by the sessile drop technique. Nevertheless, melting temperatures of either pure nickel ($T_m$=1455°C) or Ni-B alloys ($T_m$=1035-1093°C) do not allow reaching operational temperatures above 1500°C. Furthermore, these materials show a rather poor high temperature oxidation resistance. In order to overcome these limitations, we propose to evaluate a binary nickel-aluminum alloy ($Ni_{50}Al_{50}$, at%). This so called β-NiAl nickel aluminide exists as the secondary ordered solid solution over the composition range of ~45–60 at% Ni. For strictly stoichiometric composition (50/50) it melts at $T_m$=1638°C [7]. Due to a high thermal stability of a single phase, excellent oxidation resistance, high hardness and elastic modulus, low density (5.9 gcm$^{-3}$) and low price of raw materials, the NiAl nickel aluminides are being considered as candidates for many high temperature applications beyond nickel superalloys abilities [8] and few successful implementations of NiAl intermetallics have been already reached in coatings technologies [9].

In the present work, the $Ni_{50}Al_{50}$ alloy is taken into consideration as a potential brazing material for joining the $ZrB_2$ ultra-high temperature ceramic. Therefore, a specific goal of our research is to evaluate a course of high temperature interfacial phenomena (wettability, reactivity and infiltration) between molten $Ni_{50}Al_{50}$ alloy and polycrystalline zirconium diboride.

## 2. Materials and methods

The binary $Ni_{50}Al_{50}$ alloy was fabricated by the electric arc melting technique (Buehler Arc Melter MAM-1) by using properly weight mixtures of pure elements. We used the same hot pressed zirconium diboride that in our previous work [**10**].

The interfacial phenomena between molten $Ni_{50}Al_{50}$ alloy and $ZrB_2$ substrate were examined in a sessile drop experiment by using a capillary purification (CP) procedure (**Fig. 1**). In the CP procedure, the $Ni_{50}Al_{50}$ alloy was initially placed in alumina capillary located above the $ZrB_2$ substrate. After that, the experimental chamber was pumped until a vacuum of $10^{-6}$ mbar was achieved, and then a heating/cooling procedure was initiated. We used a constant heating rate of 15°Cmin$^{-1}$ to reach temperature of 1688°C ($T=1.02T_m$). At $T=500°C$ flowing argon was introduced in order to suppress the evaporation issues. After reaching the final temperature, a drop of molten alloy was squeezed through the capillary, deposited on the $ZrB_2$ substrate, held for 400 seconds and then cooled down to room temperature (at 20°Cmin$^{-1}$). Upon the experiment drop/substrate images were recorded by a high-speed camera at 100 fps. After that, the collected images were used to calculate the wetting kinetic curve (contact angle $\theta$ vs. time) by using dedicated software (Astra2, CNR-ICMATE, Italy [**11**]) and to compile the movie (**Supplementary Material 1**). More details on the applied experimental setup, are given elsewhere [**12**]. The solidified $Ni_{50}Al_{50}/ZrB_2$ couple was subjected to structural characterization by means of scanning electron microscopy coupled with X-ray energy dispersive spectroscopy (SEM/EDS) by using Hitachi TM3000 microscope and Bruker Quantax 200 analyzer.

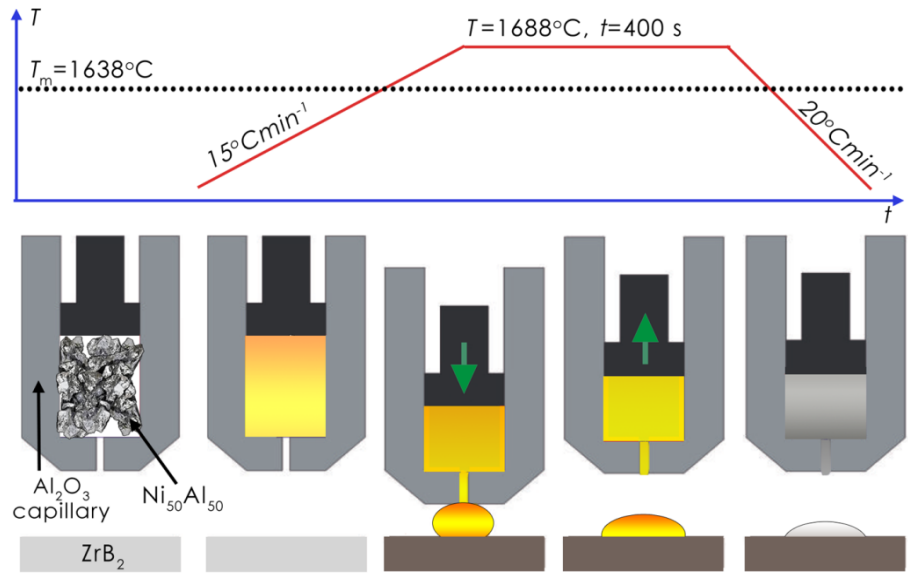

**Fig. 1.** A schematic drawing of the sessile drop method combined with capillary purification procedure used for the wettability experiment on the $Ni_{50}Al_{50}/ZrB_2$ system.

### 3. Results and discussion

The wetting kinetics recorded for the $Ni_{50}Al_{50}/ZrB_2$ couple is presented in **Fig. 2**. It is found that just after the drop deposition at 1688°C, molten $Ni_{50}Al_{50}$ alloy rapidly wets the surface of $ZrB_2$ ceramic, as it is reflected by contact angle values θ below 90°. A further holding of the molten drop results in its fast spreading over the ceramics substrate. The contact angle values decreased to θ~25° in around 180 s. In fact, we had to switch to "manual" measurements of contact angle in order to determine very low values obtained after holding time longer than 180 s. Finally, the near zero values of θ, were reached after ~250 s of the experiment. It should be noted that presently obtained contact values at $T=1.02\ T_m$ are definitely lower, while the spreading kinetic is much faster than that reported in the literature [13] for $Cu/ZrB_2$ (θ=80°), $Au/ZrB_2$ (θ=34°) and $Ag/ZrB_2$ (θ=153°) when tested at $T=1.05\ T_m$ for 35-60 minutes.

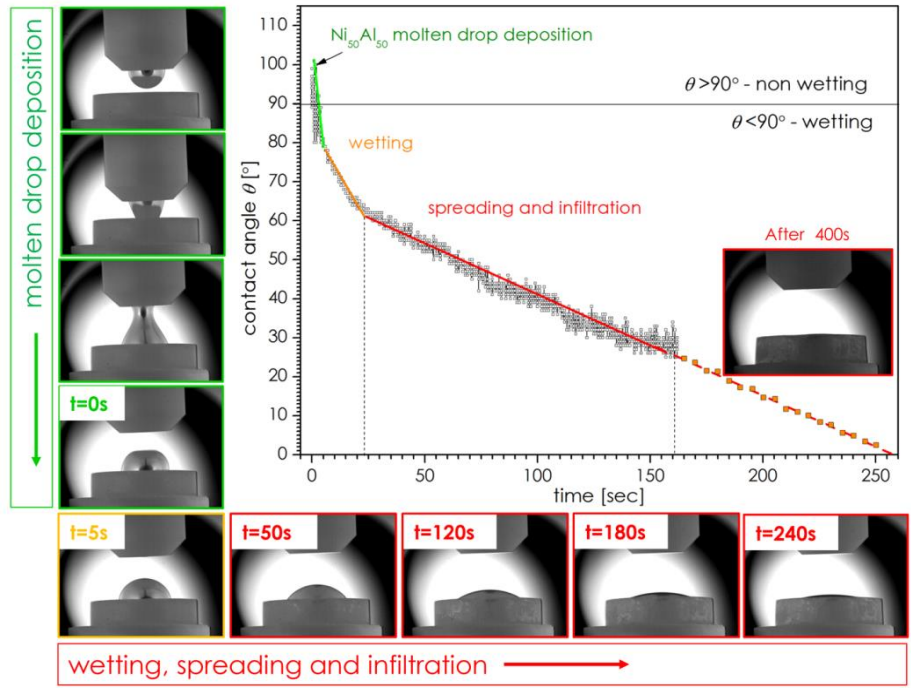

**Fig. 2.** A wetting kinetics curve calculated for the $Ni_{50}Al_{50}/ZrB_2$ system at 1688°C and corresponding recorded drop/substrate images.

The results of *post-mortem* inspections (**Fig. 3a**) showed that almost whole alloy was infiltrated into the ceramic substrate. The top view SEM/EDS analyzes performed in the vicinity of residual $Ni_{50}Al_{50}$ alloy (**Fig 3b,c**) documented the presence of Ni-enriched Ni-Al surface layer and $ZrB_2$ crystals having hexagonal morphology (**Fig. 3c**) that we have also identified by using TEM/SAED technique in our previous study on $Al/ZrB_2$ system [**10**]. The investigations carried out on the cross-sectioned couple revealed an infiltration zone having a depth of up to 2000 μm (**Fig. 4**a). The following structural features were recognized: (i) a Ni-enriched surface layer with $ZrB_2$ needled like crystals (**Fig. 4b**); and (ii) $Al_2O_3$ stringers formed along $ZrB_2$ boundaries and primary pores. Furthermore, we detected a discontinuous interfacial product layer (**Fig. 4c, d**) in which $Al_2O_3$, $ZrB_2$ and another Zr-enriched phase (most probably $ZrO_2$), were recognized.

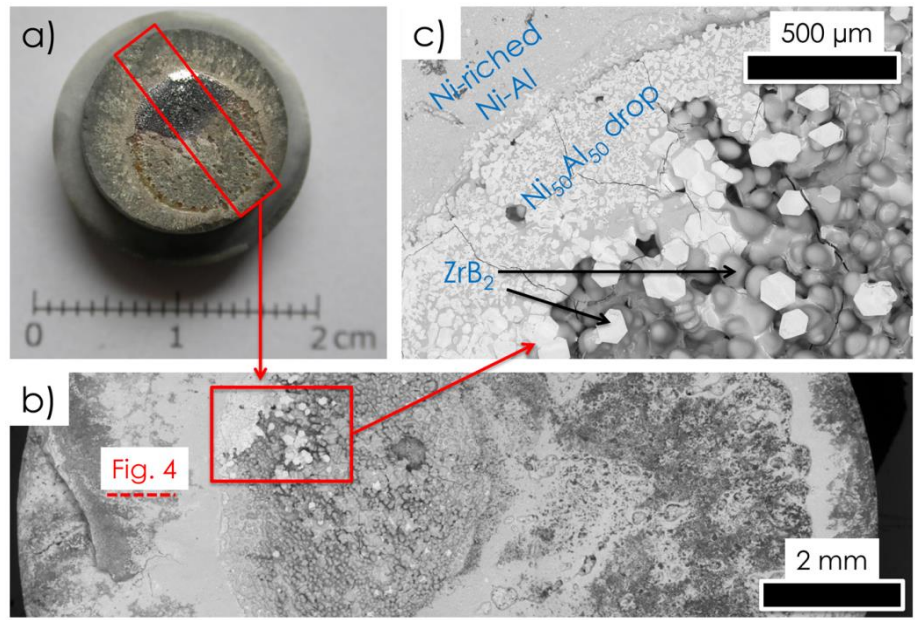

**Fig. 3.** *A macroview of the solidified Ni$_{50}$Al$_{50}$/ZrB$_2$ couple (a). The results of top-view SEM/EDS analyzes of the Ni$_{50}$Al$_{50}$/ZrB$_2$ alloy (b, c).*

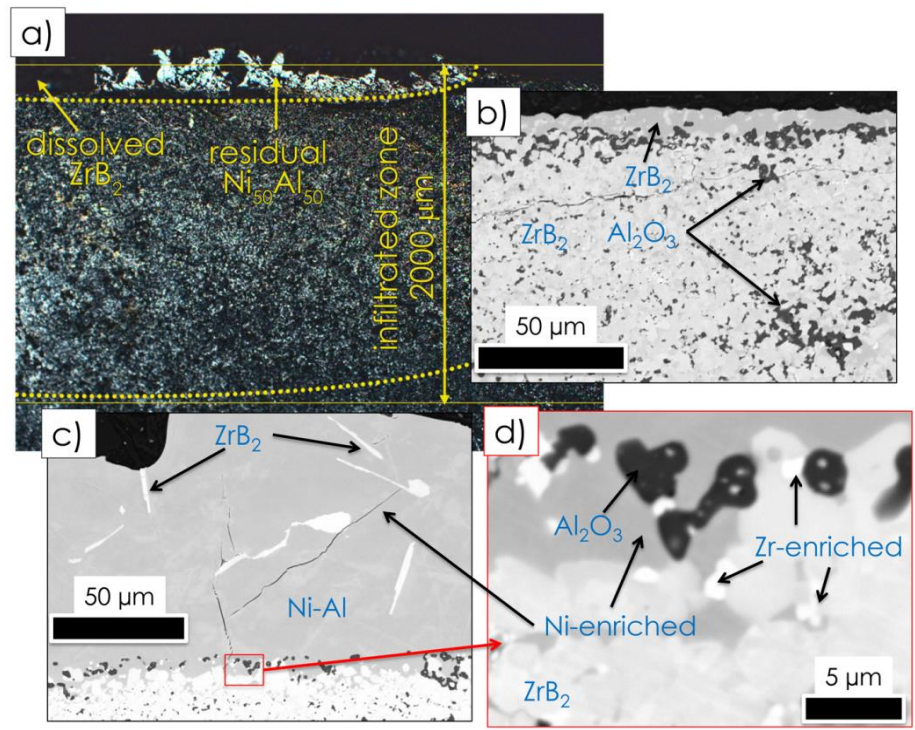

**Fig. 4.** *Low magnification light microscopy image (a) and the results of SEM/EDS studies on the cross-sectioned Ni$_{50}$Al$_{50}$/ZrB$_2$ couple showing identified structural features inside infiltration zone(a); in the residual Ni$_{50}$Al$_{50}$ alloy (b) and at the interface (c).*

## 4. Summary, conclusions and future remarks

The following conclusions are drawn regarding high temperature interfacial phenomena in $Ni_{50}Al_{50}$/ZrB2 system. The $Ni_{50}Al_{50}$ nickel aluminide shows a very good wetting with hot-sintered $ZrB_2$ ceramic at temperature of $T$=1.02 $T_m$ (1688°C). The *in-situ* observed contact angles were much lower than that reported for Cu, Ag and Au tested at respective $T$=1.05 $T_m$ temperatures. The presence of $ZrB_2$ crystals on the surface of solidified alloy and at the alloy/ceramic interface points towards a dissolution/precipitation as the main mechanism of chemical interaction. A primary porosity of $ZrB_2$ sinter (~6 vol.%) facilitates a "reactive infiltration" of the melt, that is assisted by a formation of aluminum oxide stringers along grain boundaries (a pre-oxidation of pore walls could be a possible source for oxygen). Consequently, Al is partially "consumed" and Ni-enriched layer is formed on the surface and in the vicinity of $Al_2O_3$ particles.

Although a good chemical integrity of $Ni_{50}Al_{50}$ and $ZrB_2$ has been observed in the present work, more studies are needed to prove practical usefulness of using $Ni_{50}Al_{50}$ nickel aluminide as the potential brazing material for UHTCs. In particular, experiments on fully dense $ZrB_2$ ceramic as well as some joining technological trails ought to be performed and then verified in shear tests.

**Acknowledgments**

A financial support from the Polish National Science Centre under Grant no. UMO-2012/05/D/ST8/03054 (SONATA) is gratefully acknowledged. A support of Dr Ivan Kaban (IFW Dresden) in the arc melting fabrication of $Ni_{50}Al_{50}$ alloy samples is truly appreciated.